\documentclass{iau} 

\usepackage{graphicx}
\usepackage{natbib}

\usepackage{hyperref}
\hypersetup{
    colorlinks=true,
    linkcolor=blue,
    urlcolor=blue,
    citecolor=blue,
}

\title[Imprints of Evolution on the Internal Kinematics of Globular Clusters] 
{Imprints of Evolution on the Internal Kinematics of Globular Clusters}

\author[Laura L. Watkins]   
{\href{http://orcid.org/0000-0002-1343-134X}{Laura L. Watkins}$^{1,2}$ \\
with \href{https://orcid.org/0000-0001-7827-7825}{Roeland P. van der Marel}$^3$, \href{https://orcid.org/0000-0003-3858-637X}{Andrea Bellini}$^3$,\\
\href{https://orcid.org/0000-0001-9673-7397}{Mattia Libralato}$^3$ \and \href{https://orcid.org/0000-0003-2861-3995}{Jay Anderson}$^3$}

\affiliation{
$^1$Department of Astrophysics, University of Vienna \\
T\"urkenschanzstra{\ss}e 17, 1180 Vienna, Austria \\ email: {\tt lwatkins@eso.org} \\[\affilskip]
$^2$European Southern Observatory \\
Karl-Schwarzschild-Str. 2, 85748 Garching bei M\"unchen, Germany \\[\affilskip]
$^3$Space Telescope Science Institute \\
3700 San Martin Drive, Baltimore MD 21218, USA
}

\pubyear{2019}
\volume{351}  
\jname{Star Clusters: From the Milky Way to the Early Universe}
\editors{A. Bragaglia, M.B. Davies, A. Sills \& E. Vesperini, eds.}
\begin{document}

\maketitle

\begin{abstract}
Globular clusters are collisional systems, meaning that the stars inside them interact on timescales much shorter than the age of the Universe. These frequent interactions transfer energy between stars and set up observable trends that tell the story of a cluster's evolution. This contribution focuses on what we can learn by studying velocity anisotropy and energy equipartition in globular clusters with \textit{Hubble Space Telescope} proper motions.
\keywords{astrometry, globular clusters: general, stars: kinematics, stellar dynamics}
\end{abstract}

\firstsection 
\section{Introduction}

Globular clusters are old, dense systems. The high density of stars means that the timescale for interactions between stars is short enough to be meaningful. The old age of the clusters means that the stars inside have had a long time in which to interact. Taken together, the consequence is that stars inside clusters have experienced many interactions. While any changes that result from a single interaction are small, after many interactions the cumulative effects are significant and measurable.

The history of interactions between stars in a globular cluster is imprinted on its kinematics and manifests in a number of different ways. In this contribution, I will focus only on two of these: 1) velocity anisotropy; and 2) kinetic energy equipartition.

Studying velocity anisotropy requires at least two components of motion, which necessitates proper motion data. Studying energy equipartition typically requires deep astrometry, which necessitates data from the \textit{Hubble Space Telescope} (\textit{HST}). As such, the \textit{HST} proper motion catalogues for 22 Milky Way globular clusters presented in \citet{Bellini2014} are ideally suited for this work.

\section{Anisotropy}

Velocity anisotropy is defined as the ratio between the velocity dispersions in two orthogonal directions. These can be any two orthogonal directions, and the choice of directions can depend on the structural properties of the system and the effects we wish to measure. In globular clusters, it is common to consider the projected radial and projected tangential velocity dispersions $\sigma_R$ and $\sigma_T$.

\begin{figure}
    \begin{center}
        \includegraphics[width=0.47\textwidth]{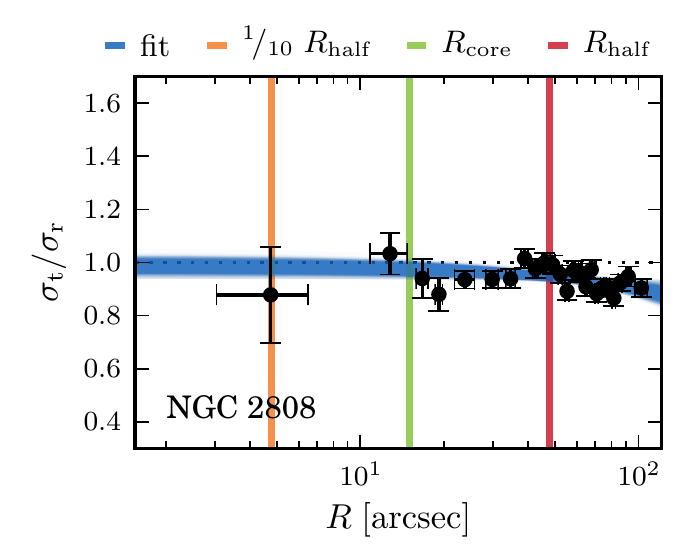}
        \qquad
        \includegraphics[width=0.47\textwidth]{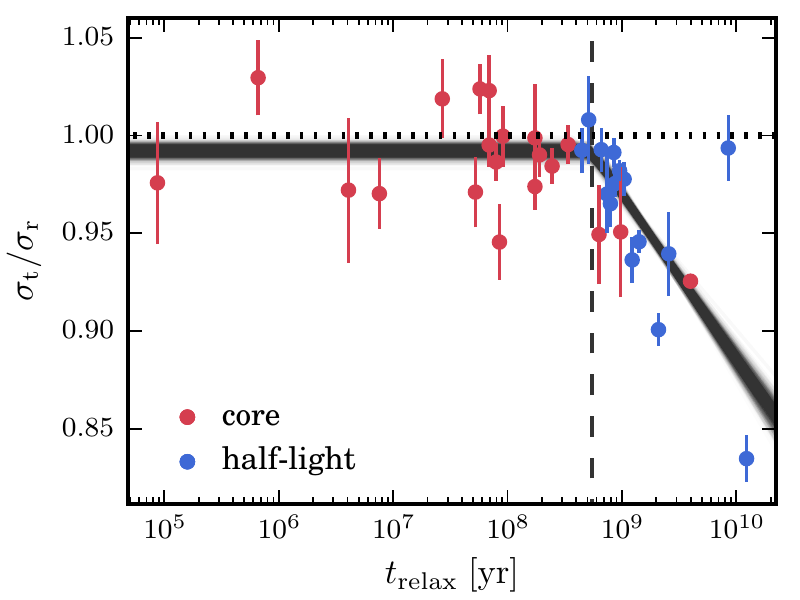}
        \caption{\textit{Left panel}: Anisotropy profile for NGC 2808. The black points show the binned profile measured with \textit{HST} proper motions. The blue lines are fits to the data; in this case we fit straight lines to the data in linear-radius, which appear curved in log-radius as shown. The red, green and orange lines show the half-light radius $R_{\rm half}$, the core radius $R_{\rm core}$, and $\frac{1}{10} R_{\rm half}$, which indicate the size of the cluster and the extent of the data. $R_{\rm half}$ and $R_{\rm core}$ were taken from \citet[][2010 edition]{Harris1996}. \textit{Right panel}: Anisotropy as a function of relaxation time for our sample of 22 clusters. Red points show the core values and the blue points show the half-light values. The grey lines show a fit that is flat for relaxation times shorter than some characteristic time and linearly decreasing at longer times. The dashed line marks the characteristic time.}
        \label{figure:anisotropy}
    \end{center}
\end{figure}

In \citet{Watkins2015a}, we measured the projected tangential-radial velocity anisotropy $\sigma_T / \sigma_R$ in all 22 GCs in the \citet{Bellini2014} sample. The anisotropy profile for a single cluster, NGC\,2808, is shown in the left panel of Figure~\ref{figure:anisotropy}. We see that the centre is isotropic ($\sigma_T / \sigma_R \sim 1$), and that the anisotropy becomes mildly radial at the outer limits of the area covered by the data ($\sigma_T / \sigma_R < 1$). In fact, we see this trend across all 22 clusters: all clusters are isotropic at their centres; some stay isotropic out to the limits of the data, and some become mildly radial in the outer regions\footnote{All of our fields are central fields, so the limits of the data are not the limits of the cluster.}. Figures 5-10 of \citet{Watkins2015a} show the corresponding profiles for the other 21 clusters in the sample.

We used the fits to estimate the anisotropy at the core and half-light radii, and compared these values to the relaxation times estimated at the core and half-light radii, from \citet[][2010 edition]{Harris1996}. These estimates are shown in the right panel of Figure~\ref{figure:anisotropy}. We see indeed that the clusters are almost all isotropic out to their core radii; some are isotropic even out to their half-light radii, while some show mild radial anisotropy at the half-light radii. The degree of anisotropy increases with relaxation time.

To quantify this, to these points we fit a simple model that assumes a constant anisotropy for relaxation times shorter than some characteristic value and a linearly decreasing anisotropy for time larger than that value. The constant anisotropy, characteristic time and slope of the decrease were all left free. We see that the best fit shows a constant anisotropy for short relaxation times very close to 1 (isotropic) and an increase in the anisotropy in the radial direction for longer relaxation times.

What can we learn about the evolution of clusters from this? We suppose that GCs in the past had some radial anisotropy then as the stars inside interact, the cluster relaxes and signals of anisotropy are washed out. The densest regions (in the cores) have the shortest relaxation times and so reach isotropy first; the intermediate regions (near the half-light radii) have longer relaxation times so have not yet reached isotropy in some clusters and we still see mild anisotropy present. The cause of the radial anisotropy, whether it was primordial or manifested later in the lifetime of the cluster, is beyond the scope of this work to say.

\section{Energy Equipartition}

When two stars pass close to one another, energy is transferred from the star with higher energy to the star with lower energy. Given enough interactions, it is expected that a stellar system will reach kinetic energy equipartition whereby all stars have the same kinetic energy. In complete equipartition we expect the velocity dispersion $\sigma$ of the stars to depend on stellar mass $m$ like $\sigma \propto m^{-\frac{1}{2}}$. In practice, most clusters will be in a state of partial equipartition with dispersions that vary with stellar mass like $\sigma \propto m^{-\eta}$ where $\eta$ quantifies the degree of equipartition (0 for none and 0.5 for full).

Naively, we might expect that given long enough clusters will reach full equipartition. However \citet{Trenti2013} used a suite of N-body simulations to predict that clusters never reach full equipartition. The simulations started with no primordial equipartition and were left to evolve. The inner regions of the clusters reached a maximum $\eta \sim 0.2$ and the degree of equipartition actually began to decrease again to level off at $\eta \sim 0.1$. The outer regions of the clusters increased steadily to $\eta \sim 0.1$ and leveled off without first overshooting. (See their Figure 3.) We would like to test this predict with real data.

Whether partial or full, equipartition means that low mass stars tend to be moving faster than high mass stars, so to measure equipartition we need to measure velocities for stars over a wide range of stellar mass. The stars for which we can most easily measure kinematics (particularly those stars that have been most frequently targeted with spectrographs to obtain line-of-sight velocities) are the bright stars above the main-sequence turn-off. Unfortunately, the phases of stellar evolution above the turn-off are fast and so the range of stellar masses covered by these stars is small and not ideal for measuring equipartition \citep[with the exception of blue stragglers stars, as discussed in][]{Baldwin2016}.

\begin{figure}
    \begin{center}
        \includegraphics[width=0.47\textwidth]{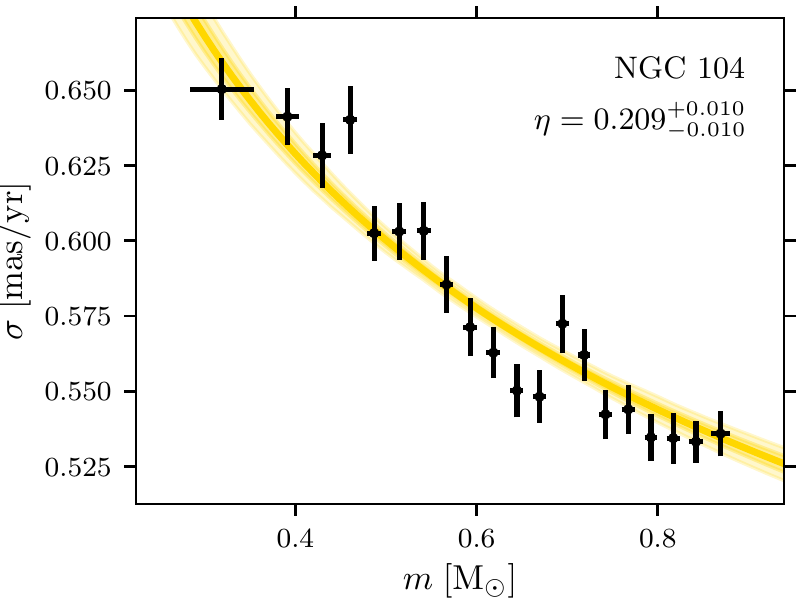}
        \qquad
        \includegraphics[width=0.47\textwidth]{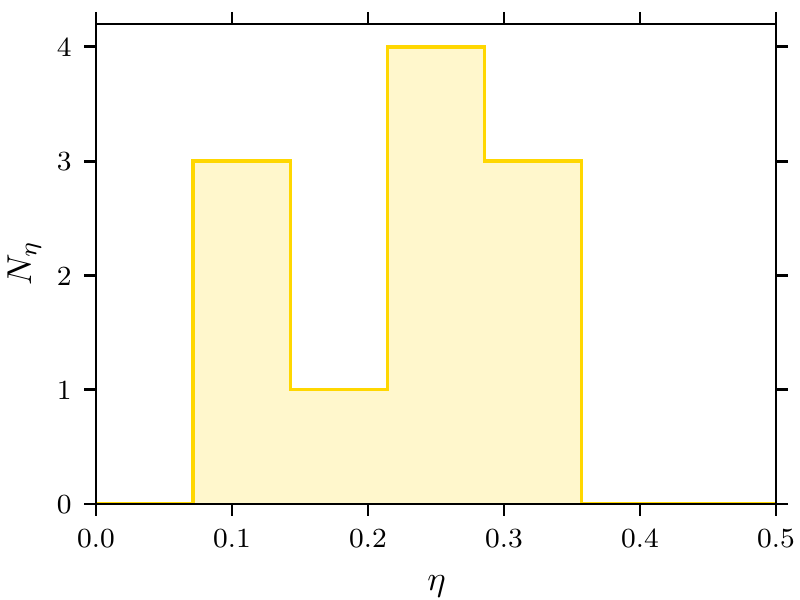}
        \caption{\textit{Left panel}: Velocity dispersion as a function of stellar mass in NGC\,104 (47\,Tuc). The black points were measured using \textit{HST} proper motions. The yellow line shows a fit to the data with the resulting $\eta$ estimate given in the top-right. \textit{Right panel}: Distribution of $\eta$ estimates for all 11 clusters for which we have performed this analysis.}
        \label{figure:etaaverage}
    \end{center}
\end{figure}

To get a decent range of stellar masses, we really need to measure velocities for stars along the main sequence. This is beyond the reach of existing spectrographs, except perhaps MUSE for the very closest clusters, and also beyond the reach of Gaia with sufficient accuracy, again except perhaps for one or two of the closest clusters. But this is achievable with \textit{HST}, although this work is still challenging and pushes both the observatory and the data analysis techniques to their limits.

In Watkins et al. (in preparation) we have measured the velocity dispersion as a function of stellar mass and estimated the degree of equipartition\footnote{This was done using only stars within the 25th and 75th percentiles in radius to mitigate the effects of incompleteness in the dense centres and edge effects in the outer regions, and the fact that velocity dispersion also changes as a function of radius.} for 11 of the 22 clusters in \citet{Bellini2014}. The left panel of Figure~\ref{figure:etaaverage} shows the results for NGC\,104 (47\,Tuc). We estimate $\eta \sim 0.2$ in this cluster, in line with the theoretical expectations. The right panel of Figure~\ref{figure:etaaverage} shows a histogram of the $\eta$ values obtain for all 11 clusters. We see that no cluster reaches full equipartition ($\eta \sim 0.5$), consistent with \citet{Trenti2013}, but we do see some values above $\eta \sim 0.2$, inconsistent with their simulations.

\begin{figure}
    \begin{center}
        \includegraphics[width=0.47\textwidth]{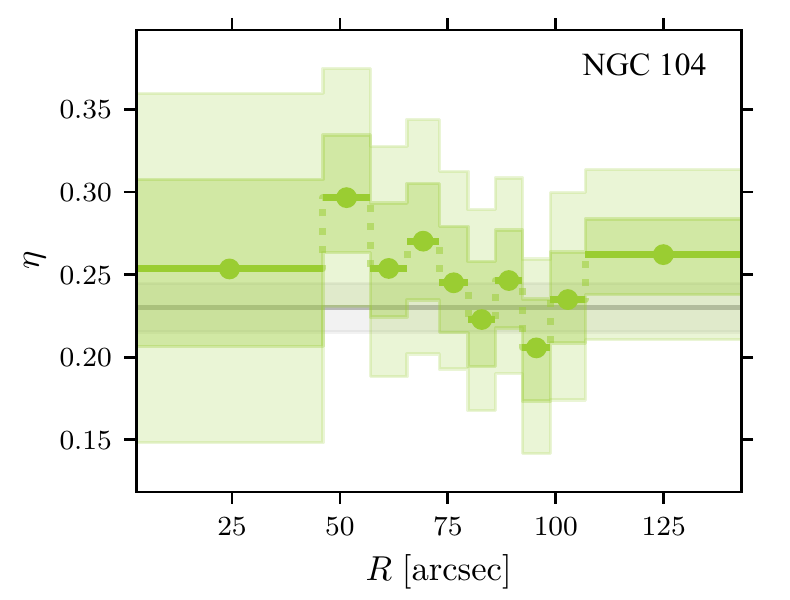}
        \qquad
        \includegraphics[width=0.47\textwidth]{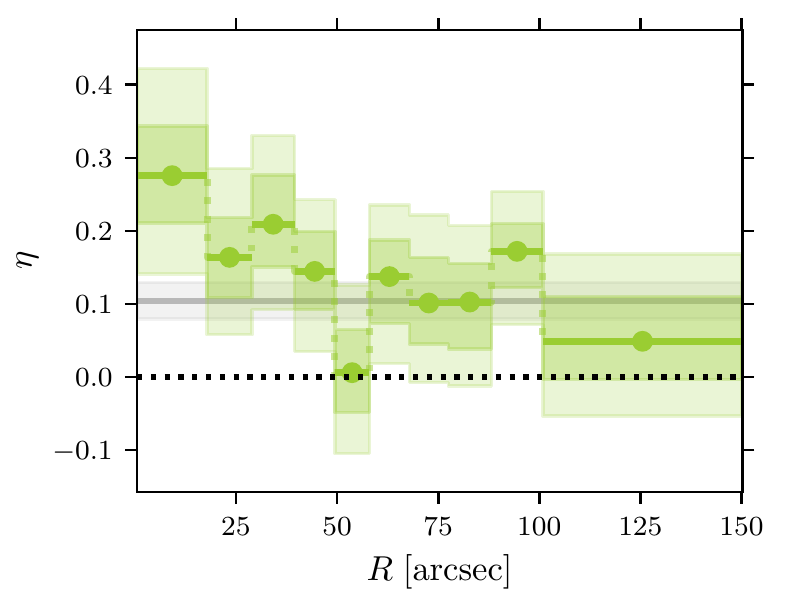}
        \caption{\textit{Left panel}: $\eta$ profile for NGC\,104 (47\,Tuc). The green lines show the median $\eta$ estimate in each radial bin; the dark shaded regions show the region spanning the 15.9 to 84.1 percentiles, and the light shaded regions the 2.5 to 97.5 percentiles. The green points mark the centre of each bin. The grey line with shaded region shows the average $\eta$ estimated in the previous analysis. \textit{Right panel}: The same for NGC\,6397. The dotted line marks $\eta = 0$, a theoretical minimum.}
        \label{figure:etaprofiles}
    \end{center}
\end{figure}

So far we have considered average $\eta$ values, however the relaxation times in the dense cluster centres are shorter than the relaxation times in the less dense regions further out. We saw the consequence of this for anisotropy in Figure~\ref{figure:anisotropy}. So the next step is to measure how $\eta$ changes as a function of radius. Figure~\ref{figure:etaprofiles} shows the $\eta$ profiles for NGC\,104 (left) and NGC\,6397 (right). NGC\,104 appears to have a flat $\eta$ profile over the range of our dataset, while NGC\,6397 shows a gradient in $\eta$ that is highest at the centre and decreases with radius, consistent with expectations.

Recently \citet{Libralato2018} studied equipartition in NGC\,362 using an improved proper motion catalogue that better controls systematics. They also saw a similar gradient in $\eta$ across the range of the data. Their smaller uncertainties highlight the efficacy of the improved data reduction and proper-motion measurement techniques.

\vspace{1mm} \footnotesize \textit{Acknowledgements}: Support for this work was provided by grants for \textit{HST} programs AR-12845 and AR-15055, provided by the Space Telescope Science Institute, which is operated by AURA, Inc., under NASA contract NAS 5-26555. LLW acknowledges support from the European Research Council (ERC) under the European Union's Horizon 2020 research and innovation programme under grant agreement No 724857 (Consolidator Grant ArcheoDyn).

\bibliographystyle{aasjournal}
\bibliography{IAUS-351}

\end{document}